\documentclass[aps,preprint]{revtex4}
\usepackage{amsfonts}
\usepackage{amsmath}
\usepackage{amssymb}
\usepackage{graphicx}
\usepackage{epstopdf}
\usepackage{epsfig}
\newcommand{\be}{\begin{equation}}
\newcommand{\ee}{\end{equation}}
\newcommand{\bea}{\begin{eqnarray}}
\newcommand{\eea}{\end{eqnarray}}
\newcommand{\ba}{\begin{array}}
\newcommand{\ea}{\end{array}}

\newcommand{\eq}[1]{(\ref{#1})}

\begin{document}


\preprint{ }
\title{Multipartite Entanglement Measures and Quantum Criticality from
 Matrix and Tensor Product States}

\vfill
\author{ Ching-Yu Huang\footnote{896410093@ntnu.edu.tw} and  Feng-Li Lin\footnote{linfengli@phy.ntnu.edu.tw}  }
\affiliation{Department of Physics, National
Taiwan Normal University, Taipei,
116, Taiwan}

\vfill
\begin{abstract}

  We compute the multipartite entanglement measures such as the global entanglement of various one- and two-dimensional quantum systems to probe the quantum criticality based on the matrix and tensor product states (MPSs/TPSs). We use infinite time-evolving block decimation (iTEBD) method to find the ground states numerically in the form of MPSs/TPSs, and then evaluate their entanglement measures by the method of tensor renormalization group (TRG). We find these entanglement measures can characterize the quantum phase transitions by their derivative discontinuity right at the critical points in all models considered here. We also comment on the scaling behaviors of the entanglement measures by the ideas of quantum state renormalization group transformations.

\end{abstract}

\maketitle

\*\\
 \section{Introduction}

Quantum phase transitions (QPTs)  are conjectured to happen in many interesting physical systems and attract much attention in modern condensed matter physics \cite{qpt}. The QPTs occur at zero temperature when a parameter is tuned to a critical value,  and are facilitated not by the thermal but the quantum fluctuations in the ground state. Especially, the correlation length diverges at the critical point and some order parameters characterize the change of the ground states across the critical point.  The diverging correlation length also implies the different parts of the system are strongly correlated quantum mechanically. Therefore, the quantum correlation and entanglement play a key role near quantum critical point. In other word, the ground state is highly entangled and the QPT should be characterized by the abrupt change of the quantum entanglement \cite{entan1, entan2}, mainly probed by the single-site von Neuman entropy (1-tangle) or the two site concurrence for one-dimensional (1D) quantum spin systems, for a review see \cite{entan3}.

   On the other hand, there are very few multipartite entanglement measures which can be used to characterize the QPTs, among which the global entanglement (GE) was proposed to probe the quantum criticality and has been tested for 1D quantum spin systems \cite{gem1,gem2,gem3,gem4}. However, as far as we know, almost nothing has been done to use GE to probe the quantum criticality of the 2D quantum spin systems, except a related fidelity computation for transverse Ising model in \cite{gemfd}. This is mainly due to the lack of the exactly solvable or numerically tractable ground states in 2D case. The situation, however, changes recently because of the new development of the generalization of the 1D  density-matrix renormalization group (DMRG) \cite{dmrg} and matrix product states (MPSs) \cite{mps,tebd,peps,itebd}  to the 2D tensor product states (TPSs) \cite{mera,ipeps,trg1,imt,trg2,trg3}.

  In this paper, we would like to probe the 1D and 2D quantum criticality numerically by the multipartite entanglement measure such as GE based on the approximation of MPSs/TPSs to the true ground states of the systems. Mainly, we will utilize the infinite time evolving block decimation (iTEBD) \cite{itebd}  numerical method to find the approximate ground states in the form of MPSs/TPSs, and then use tensor-network renormalization group (TRG) method to evaluate the entanglement measures of the corresponding TPSs.  By exploiting the translational invariance of the ground states, the TPS and TRG algorithm become quite numerically tractable \cite{trg1,imt,trg2}.

   Below we will briefly summarize the entanglement measures we will evaluate in this paper. The first is GE which  is geometrically defined as the distance between $|\psi\rangle$ and the nearest separable state $|\phi\rangle$ \cite{gem1,gem2}, i.e.,
 \be
 \label{ge} E(\psi)=-\log \Lambda_{max}^{2}\;, \qquad   \Lambda_{max}= \max |\langle\psi|\phi\rangle|
 \ee
 where the maximization of the fidelity $\Lambda$ is with respect to the variation of $|\phi\rangle$.  In the representation of TPSs, the fidelity can be computed by using TRG. Next, the TPSs with translational invariance can also be seen as the bipartite Schmidt decomposition  of the ground states, whose Schmidt number $\lambda_i$'s ($i=1,...,d_s$ with $d_s$ the physical dimension of the spin.) are the singular values obtained from the singular value decomposition (SVD) of the site tensors in performing the iTEBD. Then, we can straightforwardly evaluate the bipartite entanglement measure
\be
 S_{BP}=-\sum_i \lambda_i^2 \log \lambda_i^2\;.
\ee
Finally, for comparison we will also evaluate the single-site von Neumann entropy (1-tangle), which is denoted as $S_1$.

   Our results show that all three above entanglement measures are equally good to characterize the quantum criticality by their derivative discontinuities right at the critical point for various 1D and 2D spin systems considered in this paper. This helps to establish the universal nature of entanglement probe for quantum criticality.  Besides, we also study the scaling behaviors of the entanglement measures for 1D and 2D systems based on the idea of quantum state renormalization group (RG) transformation.  In 1D case we find the expected scaling behaviors as done for different models in \cite{entropyS1,entropyS2,entropyS3}, however, the method fails for 2D case.

    This paper is organized as follows. In the next section we will evaluate the above three entanglement measures for various 1D spin systems based on MPSs formalism, including spin 1/2 XY model and spin 1 XXZ model. In section III we will evaluate the entanglement measures for various 2D spin systems by the TPSs and TRG algorithm, including transverse Ising model, XYX model and XXZ model. In section IV we discuss the scaling behaviors of the entanglement measures for both 1D and 2D systems based on quantum state RG transformation method. We then briefly conclude in section V.


 \*\\
 \section{Entanglement measure in 1D spin system}

 For 1D quantum many-body systems, the ground states could be expressed in terms of a matrix product state (MPS) as
\begin{align}\label{mps1}
|\psi\rangle = \sum_{s_1,s_2...,s_N} Tr[A^{[1]}(s_1)A^{[2]}(s_2)...A^{[N]}(s_N)]|s_1,...,s_n\rangle
\end{align}
where $s_i=1,...,d_s$ for $i=1,...,N$, and $A^{[i]}(s_i)$ are $\chi_{i-1}$ by $\chi_{i}$ matrices, with $d_s$ denoting the physical dimension and $\chi_i$ denoting the dimension of the $i$-th bond.

 We use the iTEBD method to numerically solve for the above MPS, namely, by acting the imaginary time evolution  operator $\exp (-\tau H)$ on an initial state $|\psi_0\rangle$. The operator  $\exp (-\tau H)$ for small enough $\tau$ can be expanded through a Suzuki-Trotter decomposition as a sequence of two-site gates $U^{[i,i+1]}$. Assume translational invariance the MPS can be formally written in the following form $|\psi\rangle = \sum_{s_1,s_2...,s_N} Tr[\Gamma^A_{s_1}\Lambda^A  \Gamma_{s_2}^B\Lambda^B \cdots ]|s_1,...,s_n\rangle$, where $\Gamma_i$'s are $\chi$ by $\chi$ matrices, and $\Lambda$'s  are  $\chi$ by $\chi$ diagonal matrices of singular values. Then,  we only need to update four tensors $\Gamma^A$, $\Gamma^B$, $\lambda^A$, and $\lambda^B$ for each evolution step. In the limit $\tau\rightarrow \infty$, $\exp (-\tau H) |\psi_0\rangle$ will converge to the ground state of $H$ we are solving for. We also compute the magnetization $M_x=\langle \psi_g |\sigma_i^x |\psi_g \rangle $ and the global entanglement per site by the same method.

The global entanglement can be computed according to \eq{ge}, first we should maximize the fidelity  \eq{ge} between the quantum state $|\psi\rangle$ and a separable state $|\phi\rangle$. For a system of $N$ spin $1/2$, the separable state $|\phi\rangle = \bigotimes_{i=1}^N ( \cos(\theta_i)|0\rangle + \sin(\theta_i)|1\rangle )$ where $|0\rangle$ and $|1\rangle$ are eigenstates of $\sigma_z$.
So, the fidelity  takes the form
\begin{align}\label{matrixtrace}
  |\langle\psi|\phi\rangle|  =  |Tr(Tg^{[1]} Tg^{[2]}... Tg^{[N]} )|
\end{align}
 where $Tg^{[i]} = \sum _{s_i} A^{[i]} (s_i) \otimes B^{[i]}(s_i) $ are transfer matrices, with $B^{[i]}(0)= \cos(\theta_i),  B^{[i]}(1)= \sin(\theta_i)$, and $A^{[i]}(s_i)$ given in Eqns.\eq{mps1}. Again, assume translational invariance, $\theta_i:=\theta$ and $Tg^{[i]}:=Tg$, then the fidelity can be reduced to  $\Lambda (\theta) = |Tr ((Tg)^N)|= |\Sigma_a (\lambda_a)^N|$, where $\lambda_a$ are eigenvalues of transfer matrix $Tg$. In case the ground state is non-normalized, we should normalize  the fidelity as $ \frac{|\langle\psi|\phi\rangle|}{\sqrt{|\langle\psi|\psi\rangle|}}$ where  $ |\langle\psi|\psi\rangle|=  |Tr(Tn^{[1]} Tn^{[2]}... Tn^{[N]} )| $ with the transfer matrices $Tn^{[i]}  = \sum _{s_i} A^{[i]} (s_i) \otimes A^{[i]} (s_i)  $, or assuming translational invariance $ |\langle\psi|\psi\rangle|= |Tr (Tn^N)|= |\Sigma_e (\lambda_e)^N|$ with $\lambda_e$'s the  eigenvalues of  $Tn$.

Finally, the maximization of the fidelity $ \Lambda ({\theta})$ with respect to the angle $\theta$  can be computed by numerical algorithms, and the global entanglement follows, i.e.,
\begin{align}
  \label{nonor} E(\psi)=-\log \frac{|\langle\psi|\phi\rangle|^2}{|\langle\psi|\psi\rangle|}=-\log \frac{|\Sigma_a (\lambda_a)^N|^2}{|\Sigma_e (\lambda_e)^N|}
\end{align}

\subsection{Spin 1/2 chain}

We consider  the quantum phase transition of a spin $1/2$ quantum 1D  XY closed spin chain with transverse magnetic field. The Hamiltonian is (setting the ferromagnetic coupling to one.)
\begin{align}
 H=-\sum_{i=1}^n  \frac{1}{2} [(1+\gamma )\sigma_i^x\sigma_{i+1}^x +(1-\gamma )\sigma_i^y\sigma_{i+1}^y ]+h \sum_{i=1}^n \sigma_i^z
\end{align}
where $\sigma_i^\alpha$, $\alpha=x,y,z$ are the Pauli matrices acting on site $i$, and $h\geq0$ is  the strength of the transverse field. The model reduces to the quantum Ising model for $\gamma=1$. We will focus only on the parameter range $0< \gamma\leq 1$. The ground state for the XY spin chain can be solved analytically, which exhibits three phases: the oscillatory $(O)$, the ferromagnetic $(F)$ and the paramagnetic $(P)$ ones, and the phase changes go as follows \cite{1DXY}. As we turn on and increase  $h$, at $h= \sqrt{1-\gamma^2}$ the ground state will change from $O$ to $F$ phase if $\gamma\ne 0$. This, however cannot be captured by the order parameter $M_x$ because it is a cross-over not a phase transition. Increase $h$ further up to $h\approx 1$, we will see the transition from $F$(or $O$ if $\gamma=0$) phase with nonzero $M_x$ to $P$ one with zero $M_x$, which is a second-order quantum phase transition in the thermodynamic limit.

Here we use iTEBD method to compute the ground state wave function in the form of MPS for system size $N=2^{10}$, and the result is plotted in Fig. 1.  Besides, in Fig. 2, we plot the entanglement measure to characterize the quantum phase transitions, which reproduces the analytical results in \cite{gem2}.

\begin{figure}[ht]
\center{\epsfig{figure=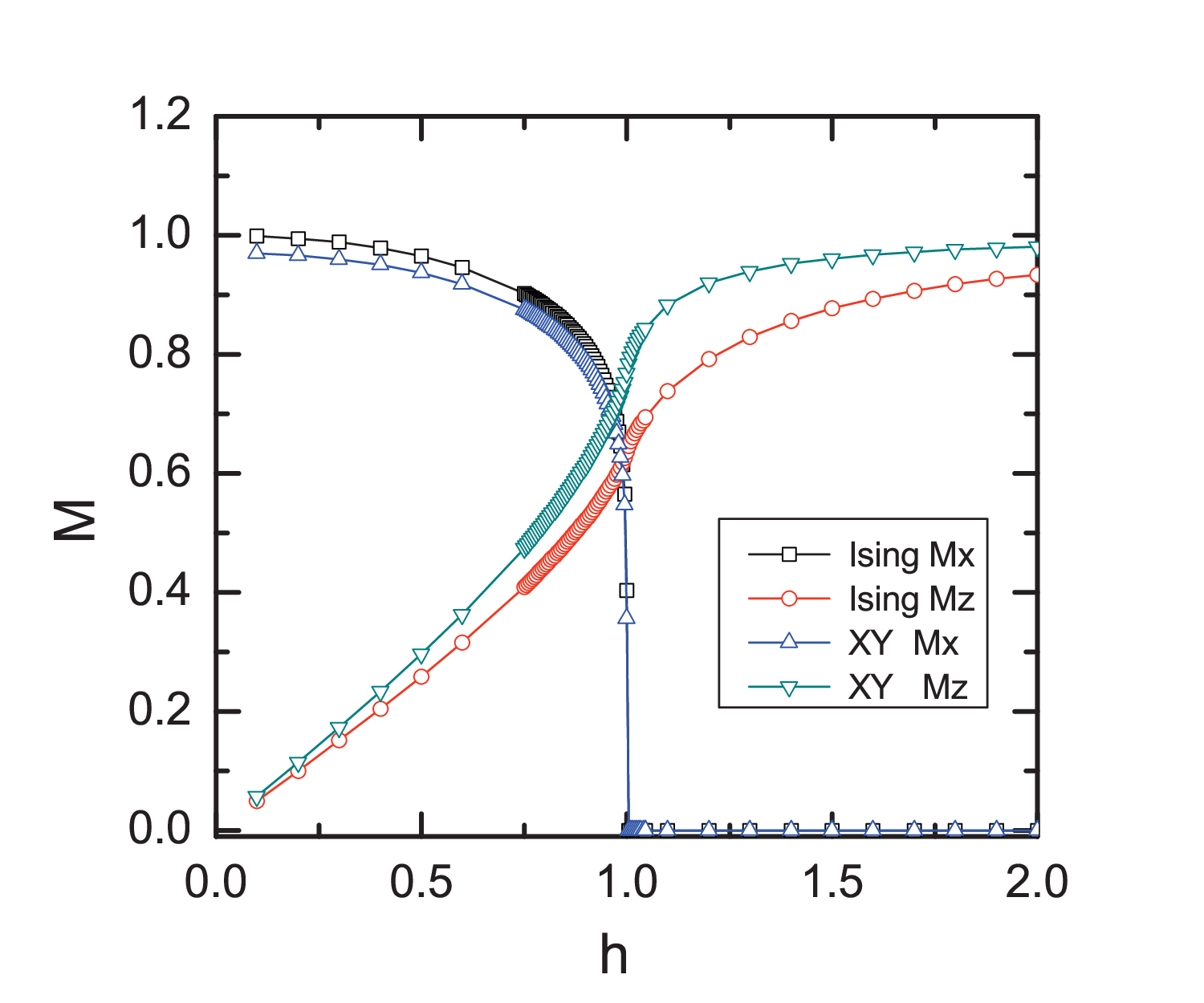,angle=0,width=7cm}}
   \caption{The magnetizations v.s. transverse magnetic field of XY spin chain using MPS with bond dimension  $\chi=16$. It indicates a 2nd-order phase transition from $F$ to $P$ phase at  $h_c\simeq 1.005$(Ising, $\gamma=0$) and $h_c\simeq 1.005$(XY, $\gamma=1/2$). However, it does not capture the phase transition from $O$ to $F$ phase.}
   \label{1DMx}
\end{figure}

\begin{figure}[ht]
\center{\epsfig{figure=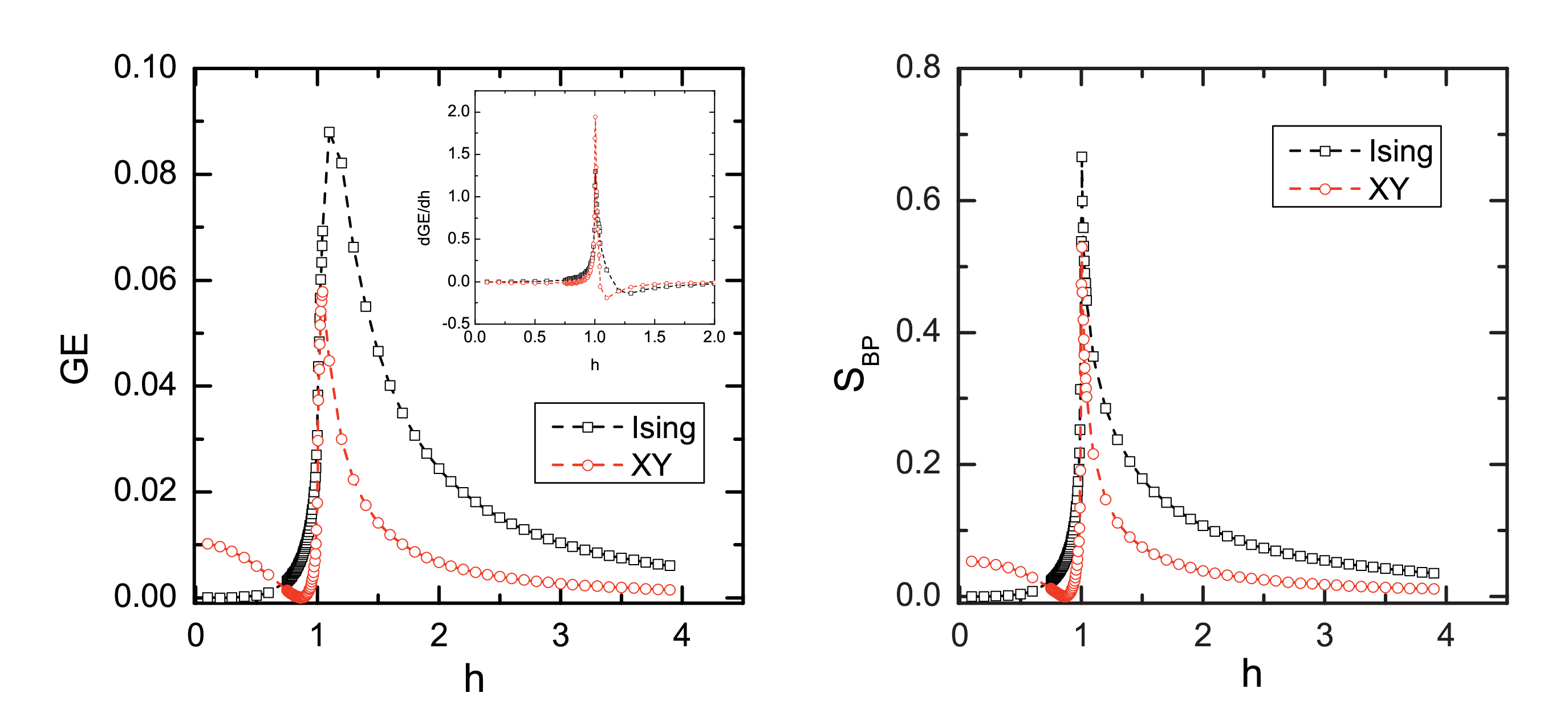,angle=0,width=15cm}}
   \caption{Entanglement measures v.s. transverse magnetic field of XY spin chain using MPS with bond dimension  $\chi=16$. (a) (Left) GE v.s. $h$. Note that near the critical point $h_c \approx 1.005$, the first derivative of GE shows $\log |h-h_c|$ scaling behavior, see the boxed graph. This agrees with the observation in \cite{gem2}. (b) (Right) $S_{BP}$ v.s. $h$.}
   \label{1DEM}
\end{figure}

In Fig. 2(a) we note that near the critical point the GE has a maximum value which implies stronger correlation near the critical point, moreover, the derivative of GE with respect to $h$ is discontinuous and diverges at $h\approx1.005$. The discontinuity of a physical quantity such as free energy or its derivatives is the notion of phase transition. Here the derivative discontinuity of the entanglement measure plays the similar role in characterizing the phase transition. On the other hand, for $\gamma=1/2$, at $h^2+\gamma^2=1$ the entanglement density vanishes and its derivative is discontinuous, it then characterizes the oscillatory (O) to ferromagnetic (F) phase.  In Fig. 2(b) the bipartite entanglement entropy $S_{BP}$ shows the same behaviors. Moreover, $S_{BP}$ is far easier to evaluate than GE since we only need the the singular values of MPS for the ground state without further numerical maneuver for calculating the expectation values.

\bigskip

\subsection{Spin 1 chain}

 To demonstrate more on the power of GE in characterizing the quantum phase transition, we now consider a spin $1$ XXZ Heisenberg  chain with anisotropy. The Hamiltonian is
\begin{align}
 H= \sum_{i=1}^n   [ S_i^x S_{i+1}^x +S_i^y S_{i+1}^y + J_z S_i^z S_{i+1}^z  ]+D \sum_{i=1}^n  {S_i^z}^2
\end{align}
where $S_i^\alpha$, $\alpha=x,y,z$ are spin-$1$ operators. The parameter $D$ represents the uniaxial  anisotropy. The ground state phase diagram of this model consists of six different phases. Here we focus our attention on phase transitions between large-$D$ phase, N$\acute{e}$el phase and Haldane phase (characterized by nonzero string order parameters) \cite{spin11,spin12,spin13}. We consider two cases of quantum phase transitions: (1) transition between large-$D$ and N$\acute{e}$el  phases by tuning $D$ but fixing $J_z$; (2)transition between N$\acute{e}$el and Haldane phases by tuning $J_z$ but fixing $D$.  We can numerically solve the ground state of the spin chain in the form of  MPS for different values of $D$ or $J_z$.

 Obviously, the ideal large-$D$ state is $|000\ldots \rangle$, and the ideal N$\acute{e}$el state is  $|1,-1,1,-1\ldots \rangle$ or $|-1,1,-1,1\ldots \rangle$. By increasing the uniaxial anisotropy $D$ the system undergoes a first-order quantum phase transition from  N$\acute{e}$el to large-$D$ phase characterized by  the staggered magnetization $M_s=\frac{1}{M}\sum_{i=1}^N (-1)^i   \langle  S_i^Z \rangle $.  Using the solved MPS, we plot in Fig. 3(a) the $M_s$ versus $D$ with fixed $J_z=2.59$ for bond dimension $\chi=4$ with system size $N=2^{10}$, and in Fig. 3(b) the $M_s$ versus $J_z$ with fixed $D=0.3$.

 \begin{figure}[ht]
\center{\epsfig{figure= 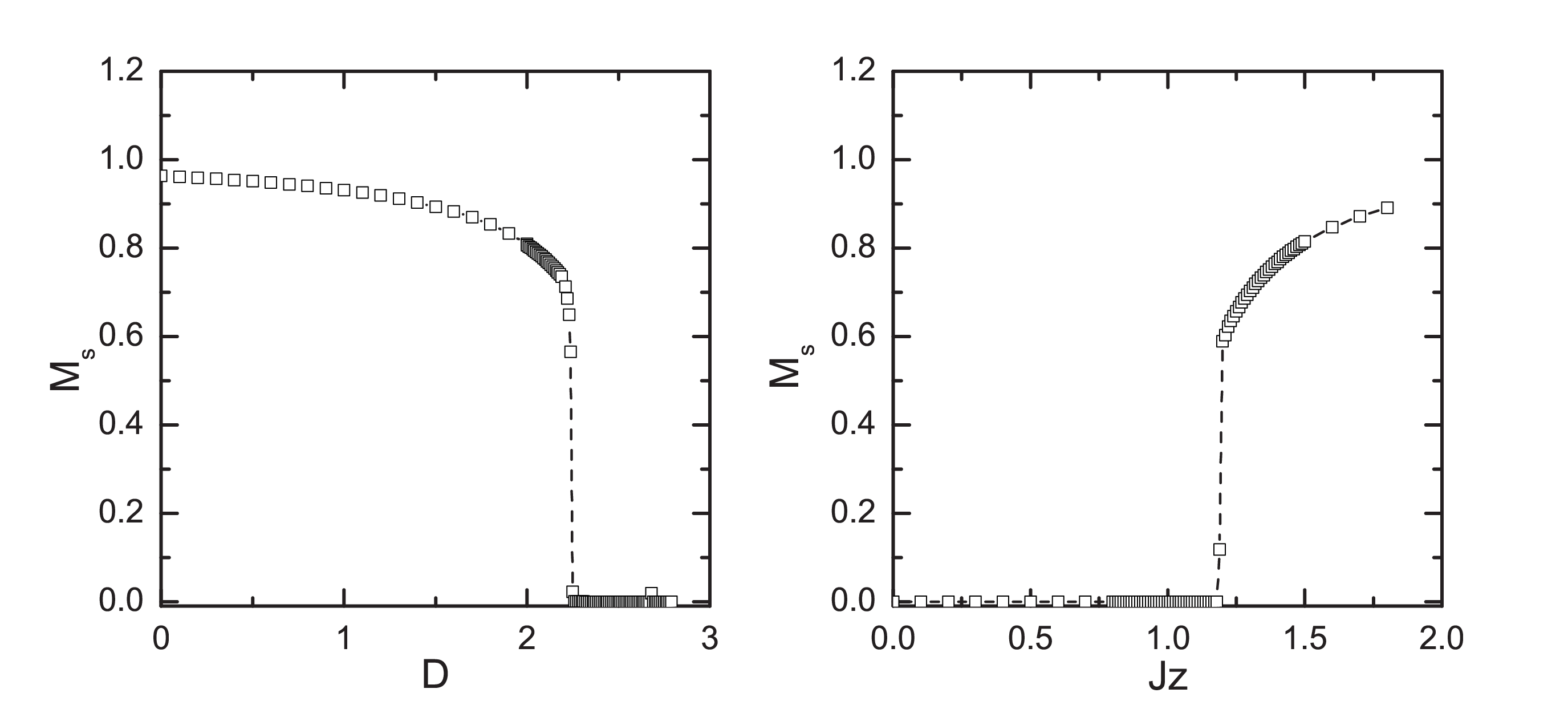,angle=0,width=14cm}}
   \caption{Staggered magnetization of spin 1 XXZ chain. (a) (Left) $M_s$ v.s. $D$ with fixed $J_z=2.59$, which indicates a phase transition from N$\acute{e}$el to large-$D$ phase at $D\simeq 2.3$. (b) (Right) $M_s$ v.s. $J_z$ with fixed $D=0.3$, which indicates a phase transition from Haldane to N$\acute{e}$el phase at $J_z \simeq 1.17$.}
   \label{spin1Ms}
\end{figure}

   On the other hand, we plot the various entanglement measures in Fig. 4, which are also capable to characterize the quantum phase transitions mentioned above.  To compute the GE for a translational invariant spin 1 chain with N$\acute{e}$el order, we should choose the following separable state ansatz
$|\phi\rangle=(|\phi_1\rangle\otimes|\phi_2\rangle)^{ \otimes \frac{N}{2}}$, where  $|\phi_i\rangle =  ( \sin\theta_i \cos\delta_i|1\rangle + \sin\theta_i \sin\delta_i|0\rangle + \cos\theta_i|-1\rangle )$ with $|1\rangle $, $|0\rangle $ and $|-1\rangle $ are the eigenstates of the $S_z$.  We can then compute the global entanglement as usual. For comparison we also calculate the 1-tangle $S_1$ and the bipartite entanglement $S_{BP}$. Again, the derivative discontinuities of the entanglement measures characterize the quantum phase transition. Moreover, we see that all three entanglement measures  do equally good job for showing the phase transition as the staggered magnetization does.

\begin{figure}[ht]
\center{\epsfig{figure=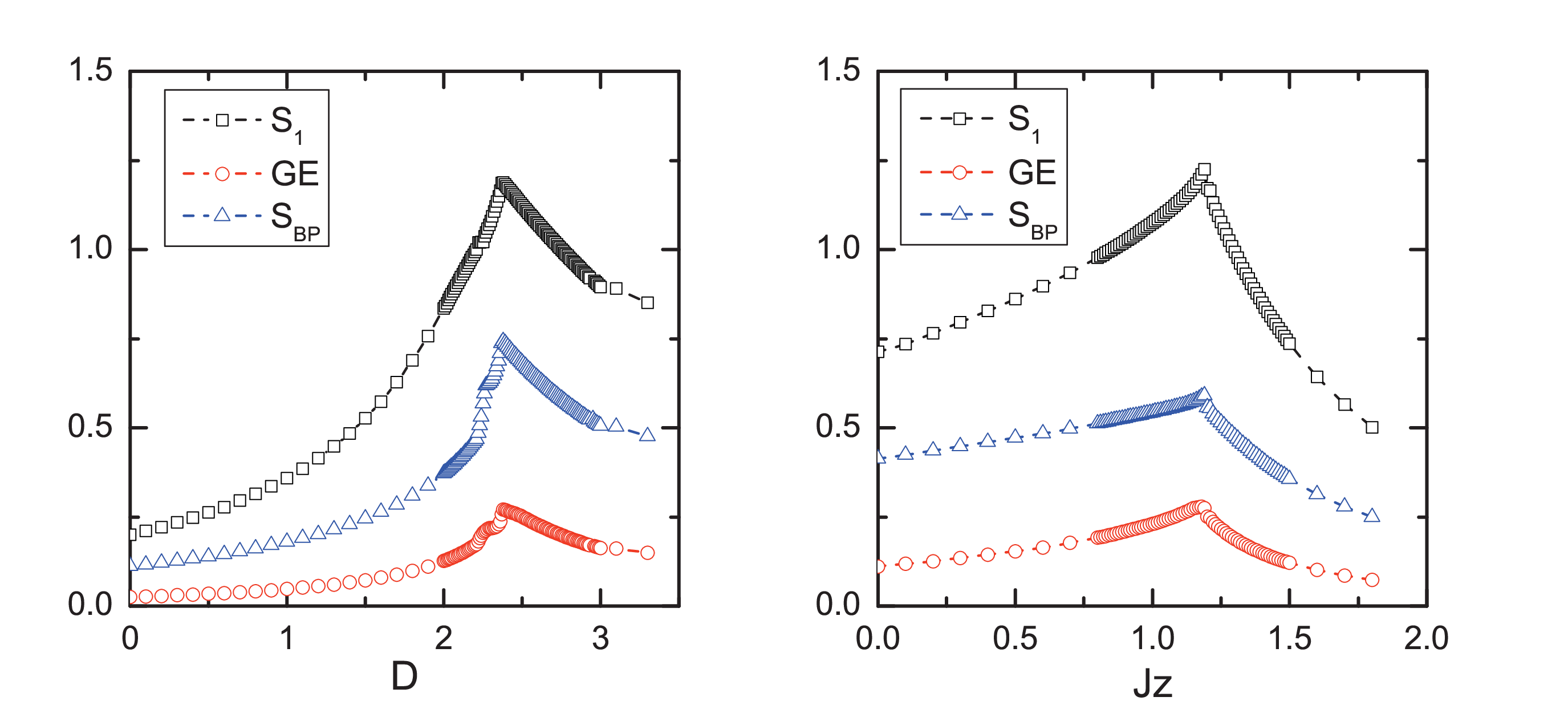,angle=0,width=14cm}}
   \caption{The entanglement measures of spin 1 XXZ chain. Their derivative discontinuity characterizes the quantum phase transition as good as the staggered magnetization does. (a) (Left) N$\acute{e}$el to large-$D$ phase; (b) (Right) Haldane to N$\acute{e}$el phase.}
   \label{spin1ent}
\end{figure}

\*\\
 \section{Entanglement measure in 2D spin systems}

 We now turn to investigate the global entanglement measure in 2D quantum spin system. Especially, we would like to see  if it is capable of characterizing the quantum phase transition or not.

 As in 1D case, we will compute the ground state wave function in the form of  the tensor product state (TPS). Several methods have been developed in the past few years, including the variational optimization of the expectation value of the energy \cite{trg2} and the iTEBD method \cite{imt}. The ground states of 2D lattice systems obtained by using any of the above methods, can be expressed in terms of a TPS as
\begin{align}
|\psi\rangle = \sum_{s_{1,1},s_{1,2},\ldots ,s_{N,N}} tTr[A^{[1,1]}(s_{1,1})A^{[1,2]}(s_{1,2})\ldots A^{[N,N]}(s_{N,N})]|s_{1,1},s_{1,2},\ldots ,s_{N,N}\rangle
\end{align}
where the values of the physical spin $s_{i,j}=1,\cdots d_s$  for $i,j=1,\ldots , N$ ;  $tTr$  is to sum over all indices of tensors; and $A^{[i,j]}_{r,d,l,u}(s_{i,j})$'s are rank-five tensors with the bond indices $r,d,l,u=1,\cdots,\chi$. We again call $\chi$ the bond dimension and $d_s$ the physical dimension. In the following, we will use the iTEBD method to calculate the TPS on square lattice, which is quite similar to the 1D case though we should update the tensors and the Schmidt values by more subtle singular value decompositions, for some example see \cite{2DXXZ3}. In order to minimize the Trotter
error of iTEBD, we usually take $\delta \tau=0.1$.

   To compute GE, we need the separable state ansatz as $|\phi\rangle = \bigotimes_{i,j=1}^N ( \cos(\theta_{i,j})|0\rangle + \sin(\theta_{i,j})|1\rangle )$. Then, the norm of a TPS and its overlap with a separable state are  $ |\langle\psi|\psi\rangle|=|tTr(Tn^{[1,1]}\ldots  Tn^{[N,N]} )|  $  and   $ |\langle\psi|\phi\rangle|=  |tTr(Tg^{[1,1]}\ldots Tg^{[N,N]} )| $ respectively, where $Tn^{[i,j]}_{(rr^\prime,dd^\prime,ll^\prime,uu^\prime)} = \sum _{s_{i,j}} A^{[i,j]}_{r,d,l,u} (s_{i,j}) \otimes A^{[i,j]}_{r^\prime,d^\prime,l^\prime,u^\prime} (s_{i,j})  $ and $Tg^{[i,j]}_{(r,d,l,u)} = \sum _{s_{i,j}} A^{[i,j]}_{r,d,l,u} (s_{i,j}) \otimes B^{[i,j]}(s_{i,j}) $ are  the transfer matrices, and $B^{[i,j]}(0):= \cos(\theta_{i,j}),  B^{[i,j]}(1):= \sin(\theta_{i,j})$.  For a translationally invariant system the fidelity $\Lambda= \frac{|\langle\psi|\phi\rangle|}{\sqrt{|\langle\psi|\psi\rangle|}}$  can be expressed as
\begin{align}
\Lambda = \frac{|tTr (Tg\otimes Tg\otimes Tg \otimes \ldots)|}{\sqrt{ |tTr (Tn \otimes Tn \otimes Tn\otimes \ldots)|}}
\end{align}
where  $Tg^{[i,j]} \equiv Tg$, $Tn^{[i,j]} \equiv Tn$, $\forall i,j$. We can then obtain the GE by vary it with respect to $\theta_{i,j}$.

  Unlike the 1D case where the matrix trace in \eq{matrixtrace} is computationally tractable. In 2D, the tensor trace  $tTr$ over all connected indices of the above norm or overlap is exponentially hard. Instead, a method called tensor renormalization group (TRG) is developed in \cite{trg1,trg2,imt,trg3} to make the double tensor trace polynomially calculable by merging the sites and truncating the bond dimension of the merging lattices according to the relevance of the components in the Schmidt decomposition.  The cutoff of the merging bond dimension is denoted as $D_{cut}$ which controls the accuracy of the computation. Each step of the TRG will reduce the number of sites by half. Eventually, the tensor-network is reduced to only four sites and the double tensor trace for the norm or overlap can be calculated easily. For more detailed description of the TRG method, please see  \cite{trg1,trg2,trg3}. We will then adopt the TRG method to numerically evaluate the entanglement measure such as GE for the ground state in the form of TPS on the 2D square spin lattice.

 \*\\

 \subsection{The 2D transverse  Ising model}

 To calibrate the accuracy of our numerical codes, we first consider the 2D spin $1/2$ transverse Ising model on a square lattice, which is well-studied using TPS and TRG, for example see \cite{trg2}. As a by-product we find that the entanglement measure such as GE can also characterize the quantum phase transition of the model.
The Hamiltonian is
\begin{align}
H=\sum_{<i,j>} \sigma_i^z\sigma_{j}^z + h \sum_i \sigma_i^x \;,
\end{align}
where $<i,j>$ stands for the nearest neighboring sites.

\begin{figure}[ht]
\center{\epsfig{figure=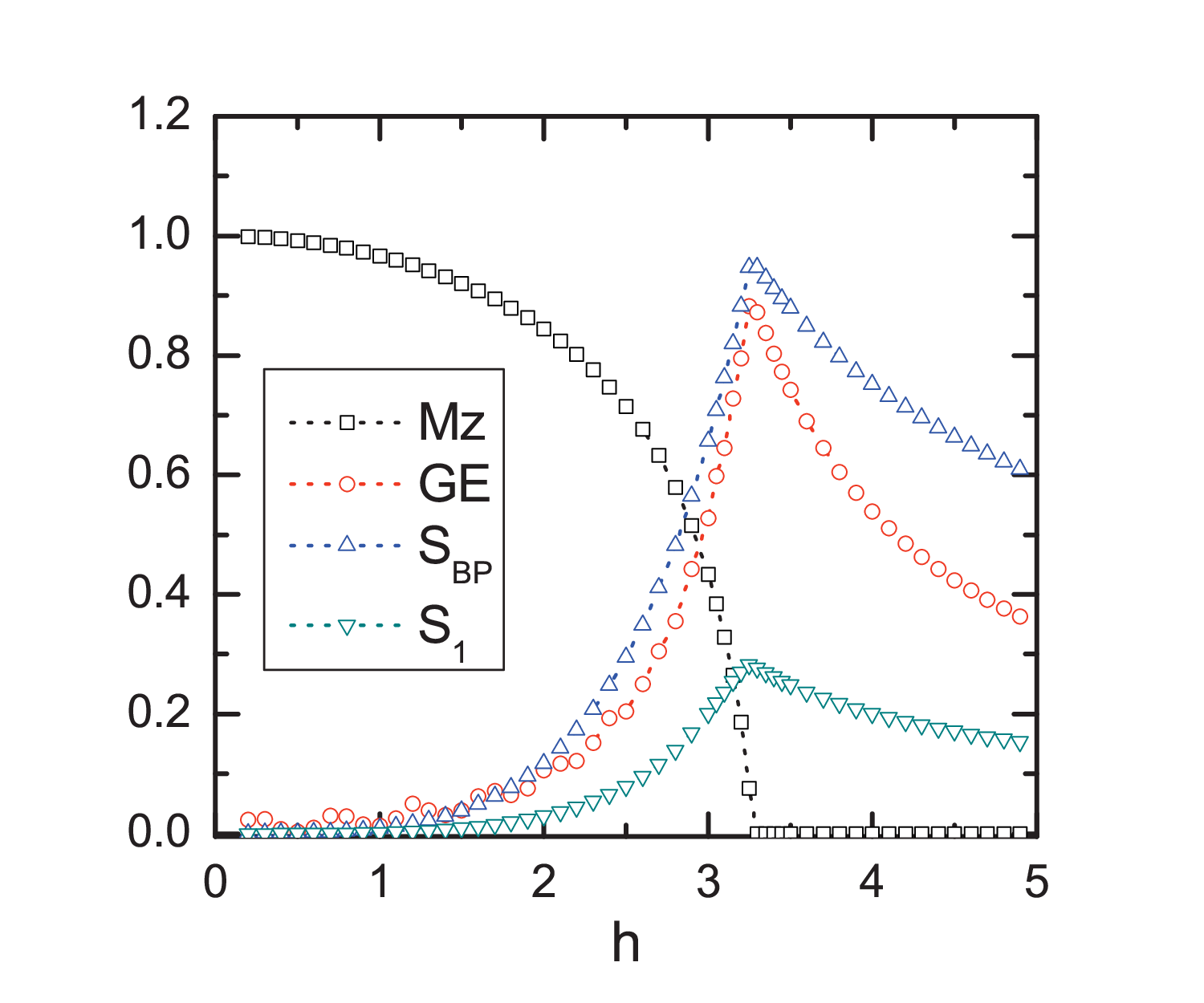,angle=0,width=8cm}}
   \caption{The 2D transverse Ising model:  $M_z$, GE, $S_{BP}$ and $S_1$ v.s. the transverse magnetic field $h$ for a system of size $2^5\times 2^5$. We adopt TPS and TRG with $\chi=4$ and $D_{cut}=16$. The derivatives of these quantities all have a discontinuity at $h\approx 3.25$. In this figure, the GE and $S_{BP}$ are scaled up by a factor of $25$ and $10$, respectively. }
   \label{2Dising}
\end{figure}

By tuning $h$, there is a quantum phase transition characterized by the $Z_2$-symmetry order parameter $M_z= \langle\psi|\sigma_i^z|\psi\rangle$, and its occurs at $h\approx 3.04$ obtained from the unbiased quantum Monte Carlo simulation \cite{2DIsing1}.  Our result using TPS and TRG is plotted in Fig. 7. We consider the bond dimension $\chi=4$ for a system size $2^5\times 2^5$, and keep $D_{cut}=16$ to ensure the accuracy of the TRG calculation. We find a second order phase transition at $h \simeq 3.25$, which can be characterized either by the order parameter or the entanglement measures.  Note that we obtain the bipartite entanglement per length  $S_{BP}$ by averaging $-\sum {\lambda_{i,j}}^2\log{\lambda_{i,j}}^2$ over eight bonds of a plaquette.

 \subsection{The 2D XYX model}

  We now consider the 2D anti-ferromagnetic spin $1/2$ XYX model in an uniform z-axis external magnetic field $h$, i.e.,
\begin{align}
 H= \sum _{<i,j>} (\sigma_i^x \sigma_j^x+\Delta_y \sigma_i^y \sigma_j^y+ \sigma_i^z \sigma_j^z)+h \sum_i  \sigma_i^z.
\end{align}
For $\Delta_y<1$  the system has easy-plane(EP) behavior. The order parameter in the EP is the x-axis magnetization $M_x$ by tuning $h$ \cite{2DXYX1,2DXYX2}.

 The model shows two quantum critical points, and only one can be captured by the order parameter $M_x$. This is similar to the phase diagram of 1D XY quantum spin chain. Before turning on $h$, the ground state is in N$\acute{e}$el phase with nonzero $M_x$. As $h$ increases to some critical value $h_f$ called factorizing field value (its theoretical value is $h_f=2\sqrt{2(1+\Delta_y)}$ \cite{2DXYX1,2DXYX2}), the ground state becomes separable product state but still with nonzero $M_x$. Therefore, the order parameter cannot characterize the cross-over but the entanglement measures do.  Increasing $h$ further to the another critical value $h_c$, there is a second-order phase transition from N$\acute{e}$el to the disorder phase with zero $M_x$. We also find that the entanglement measure can characterize this quantum critical point by their derivative discontinuities. Finally, as $h\rightarrow \infty$, the system is in a polarized state with the magnetization of z-direction saturated to the unity.

  As in 1D XY spin chain case, there is again a subtlety in choosing the separable state ansatz for evaluating GE in the N$\acute{e}$el phase. By N$\acute{e}$el and translational symmetry, we should parameterize  the separable states via four different states, i.e.,  $|\phi\rangle=(|\phi_1\rangle\otimes|\phi_2\rangle \otimes |\phi_3\rangle\otimes|\phi_4\rangle)^{ \otimes \frac{N}{4}}$ with  $|\phi _i\rangle = \cos(\theta_{i})|0\rangle + \sin(\theta_{i})|1\rangle $.    We then compute the ground state and its entanglement measures by using iTEBD for TPS and TRG with $\chi=4$ and $D_{cut}=16$.

We  present our numerical results in Fig. 6 with $\Delta_y = 0.25$ and systems size of $2^5\times 2^5$,  which shows a second-order quantum critical point at $h_c\simeq 3.5$ characterized by both the order parameter and the entanglement measures, and also the factorizing field value at $h_f\simeq 3.16$ at which the entanglement measures vanish as expected.  Our results agree with the previous one based on 1-tangle \cite{2DXYX2}.

\begin{figure}[ht]
\center{\epsfig{figure=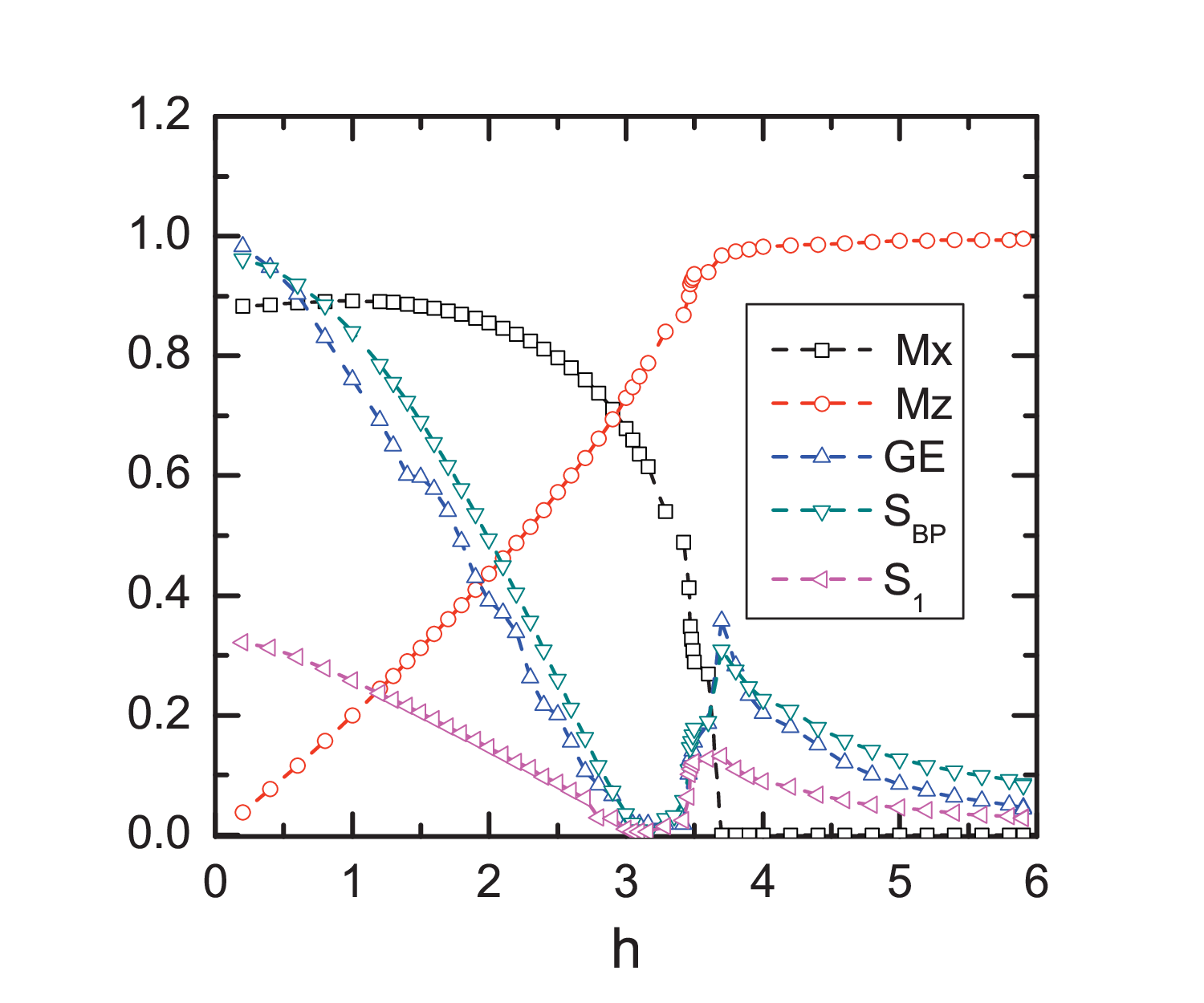,angle=0,width=12cm}}
   \caption{The 2D XYX model with $\Delta_y=0.25$: $M_x$, $M_z$, GE, $S_{BP}$ and $S_1$ v.s. the transverse magnetic field $h$ for a system of size $2^5\times 2^5$. We adopt TPS and TRG with $\chi=4$ and $D_{cut}=16$. The derivative of these quantities shows a discontinuity at $h\approx 3.5$, and the entanglement measures vanish at the factorizing field value $h\simeq 3.16$. In this figure, the GE and $S_{BP}$ are scaled up by a factor of $25$ and $10$, respectively. }
   \label{2DXYX}
\end{figure}

 \subsection{The 2D XXZ model}

 Finally, we consider the 2D spin $1/2$ XZX model in an uniform z-axis external magnetic field $h$, i.e.,
\begin{align}
 H=-\sum _{<i,j>} (\sigma_i^x \sigma_j^x +  \sigma_i^y \sigma_j^y - \Delta \sigma_i^z \sigma_j^z)-  h \sum_i  \sigma_i^z.
\end{align}
 In the large spin-anisotropy limit $( \Delta> 1)$, it has been shown that \cite{2DXXZ1,2DXXZ2,2DXXZ3} as $h$ being increased, a first-order spin-flop quantum phase transition from N$\acute{e}$el to spin-flipping phase  occurs at some critical field $h_c$. As we further increase $h$ to another critical value $h_s = 2(1+\Delta)$, the fully polarized state is reached.

 \begin{figure}[ht]
\center{\epsfig{figure=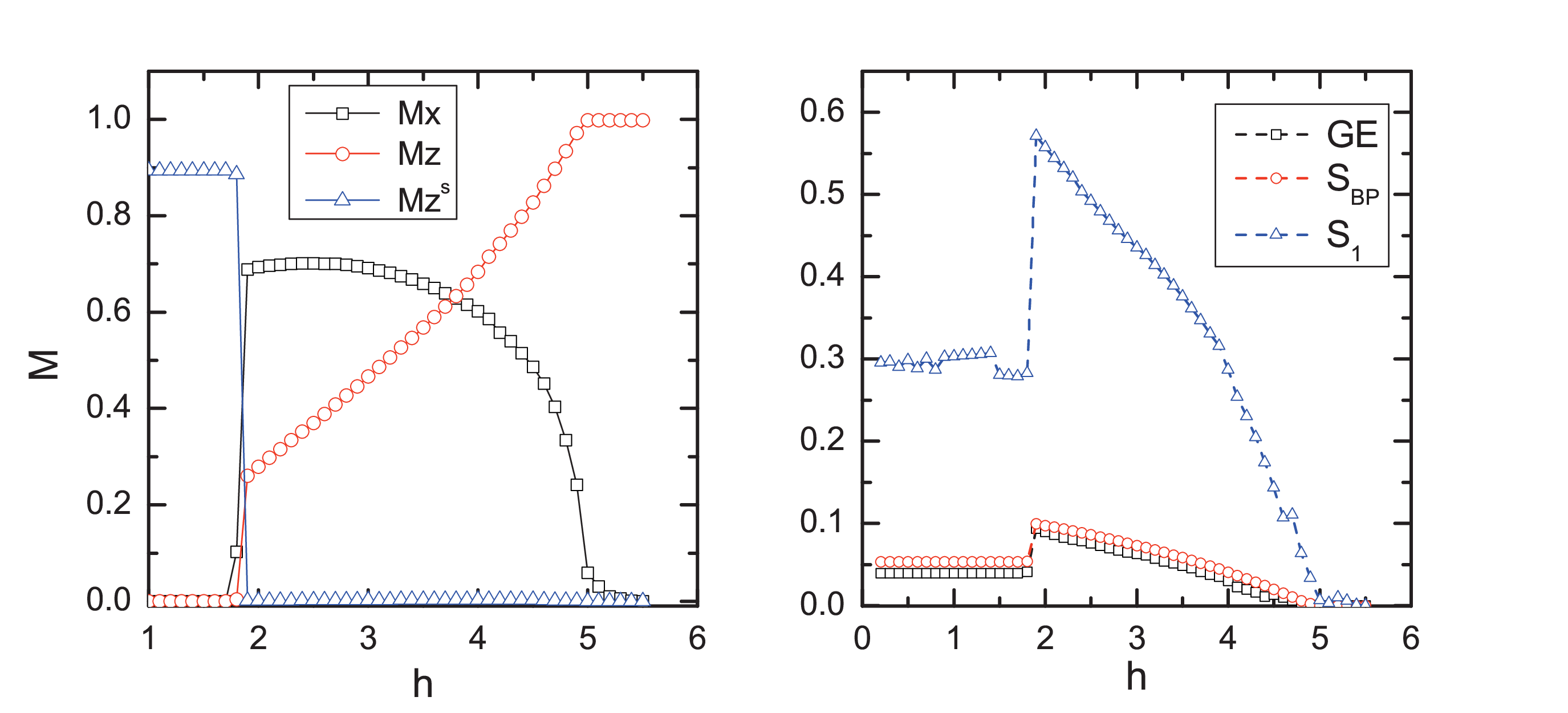,angle=0,width=16cm}}
   \caption{The 2D XXZ model: (a) (Left) Magnetizations v.s. the transverse magnetic field $h$. The derivative of all three magnetizations has a discontinuity at $h_c\simeq 1.8$ for spin-flop phase transition, and only the $M_x$ and $M_z$ show discontinuity at $h_s\simeq 5$. (b) (Right) Entanglement measures v.s. $h$. Their derivative discontinuities characterize both quantum critical points. Here we adopt TPS and TRG with $\chi=4$ and $D_{cut}=16$.  }
   \label{2DXXZ}
\end{figure}

    Here we consider the case with $\Delta = 1.5$.  Our results are plotted in Fig. 7 from solving ground state via iTEBD and TPS, and evaluating the order parameter and the entanglement measures. In Fig. 7(a) we plot the expectation values of the z-direction staggered magnetization $M_{z}^s$, the uniform one $M_{z}$, and the x-direction uniform magnetization $M_x$.
It shows that  the N$\acute{e}$el order $M_{z}^s$, will suddenly drop to zero at $h_c\simeq 1.8$ and remain zero after that. On the other hand, $M_{z}$ is zero  for $h<h_c$ but suddenly start to grow at $h=h_c$, and finally reaches a saturated value at $h_s\simeq 5$. Similarly, $M_x$ is nonzero only for $h_c<h<h_s$. Both $M_x $ and $M_z$  can be used to characterize the quantum phase transition from spin-flipping to the fully polarized state. Therefore, not all three order parameters can be used to characterize both quantum critical points. On the other hand, in Fig. 7(b) we can use either 1-tangle, GE or $S_{BP}$ to characterize both quantum critical points by their derivative discontinuities. This suggests that all the entanglement measures discussed above may belong to the same universal class. This seemingly universal feature can be seen as an advantage to use entanglement measure as an index for quantum phase transition. Finally, unlike the usual derivative discontinuity of the entanglement measure for the phase transitions encountered before, we note that the first order spin-flop transition is characterized by the more dramatic change, namely, by the sudden jump of both order parameter and entanglement measures. This indicates that the degree of discontinuity of the entanglement measures as well as order parameter may be related to the order of phase transition.

 \*\\
\section{Comment on scaling behavior of entanglement measure}

      In the usual renormalization group (RG) sense as the coarse graining, the quantum critical points are usually the IR fixed points of the original spin system, and around which the correlation length diverges and some scaling symmetries emerge. Theses scaling symmetries then dictate the scaling law of some physical quantities. On the other hand, it was known that the von Neumann entanglement entropy obeys the area law \cite{Holzhey:1994we}, and especially in 1D the entanglement entropy obeys the log scaling of the block size at critical point, with the coefficient proportional to the central charge of the underlying $1+1$ conformal field theory (CFT), otherwise it will saturate to a constant value \cite{entropyS1,entropyS2}.

     We may then wonder if the more quantum information motivated entanglement measures such as GE will obey the similar scaling law or area law or not. However, the straightforward way of computing block entanglement measure is quite difficult because the physical dimensions of the block will grow with the block size exponentially, and make the numerical calculation formidable.  Fortunately, in the framework of MPS or TPS one can adopt the method of quantum state RG transformation to merge the lattice sites \cite{qsrg}, by which the block entanglement measure of size $L$ is the same as the single-site entanglement measure after $L$ steps of quantum state RG transformation \cite{entropyS3,entropyS4}.  Moreover, at each step of quantum state RG transformation, the physical dimension is bounded by $\chi^2$ for 1D and $\chi^4$ for 2D. However, this kind of truncation of physical dimension only throws away the local informations but still keeps the non-local ones intact. In this way, we can compute the block entanglement measure for MPSs/TPSs.

     In the following we will comment on the scaling behaviors of entanglement measure of MPSs/TPSs in 1D and 2D quantum spin systems.

\subsection{1D scaling via MPSs}

      When discussing the scaling behaviors of the block entanglement measure near the quantum critical point based on MPSs/TPSs approximation to the true ground state of the spin systems, we should be aware that the bond dimension $\chi$ will play the role of the effective correlation length $\xi_{\chi}$ of the MPSs/TPSs, which will not diverge at the critical point as for the correlation length of the true ground state $\xi$.  The difference between $\xi_{\chi}$ and $\xi$ is because we approximate the true ground state by the MPSs/TPSs with finite $\chi$. Therefore, $\xi_{\chi}$ should be related to the bond dimension $\chi$. Indeed, in \cite{entropyS3} the authors found the following scaling law for the 1D Ising model,
\be\label{chis}
 \xi_{\chi} \simeq \chi^2\;,
\ee
and the block von Neuman entropy (1-tangle) will be bounded by
\be\label{SLb}
S_{\chi} \simeq {c\over 6} \log \xi_{\chi} \simeq {c\over 3} \log \chi
\ee
where $c$ is the central charge of the corresponding IR CFT.  Therefore, the typical scaling law of the block von Neumann entanglement $S_{L}\simeq {c\over 6} \log L$  \cite{Holzhey:1994we,gemrg1} will saturate to the value of \eq{SLb} as $L>\xi_{\chi}\simeq \chi^2$. On the other hand, when away from the critical point, the correlation length $\xi$ could be far smaller than $\xi_{\chi}$ and the MPSs/TPSs approximate the true ground state well.

\begin{figure}[ht]
\center{\epsfig{figure=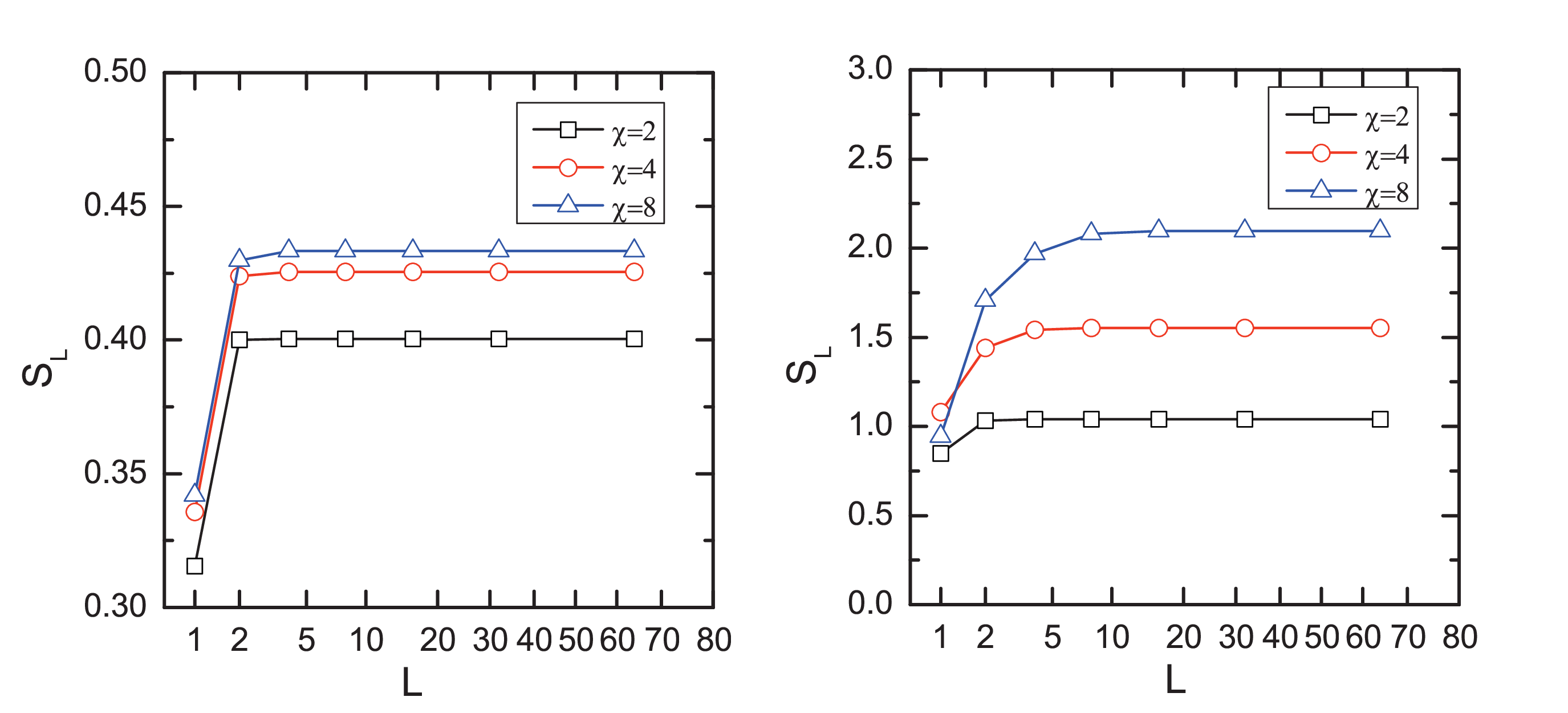,angle=0,width=16cm}}
   \caption{The block entanglement entropy $S_L$ for spin 1 XXZ chain with the fixed $J_z=2.59$ and the quantum critical point at $D\simeq 2.3$.  We adopt the MPSs of $\chi=2,4,8$ for the ground state. (Left) Off the critical point, the block entanglement entropy saturates around $L=4$. (Right) Very near the critical point, the block entanglement entropy saturates around $L\simeq\chi^2$, which can be seen more clearly from the numerical data.}
   \label{Table1}
\end{figure}

   Based on the above discussion in the framework of the MPS/TPS, one may wonder if the scaling law \eq{chis} and the entropy bound \eq{SLb} are universal or not for different 1D spin systems and various entanglement measures. In Fig. 8 we plot the block von Neumann entanglement entropy of size $L$ for the 1D spin 1 XXZ chain (central charge $c=1$ for the corresponding IR CFT.), and the results support the scaling law \eq{chis} and the entropy bound \eq{SLb}.

   As mentioned above, the straightforward calculation of block entanglement entropy is formidable because of the exponential growth of the block's physical dimension. Instead, we follow the method of quantum state RG transformation proposed in \cite{qsrg} so that the physical dimension is capped by $\chi^2$, i.e., the correlation length of the MPS, and then we can obtain the block entanglement entropy of size $L$ by performing $L$ steps of quantum state RG transformation. However, we will see that the quantum state RG transformation method fails to get the expected area law for the block entanglement entropy in 2D case due the exponential growth of bond dimension when merging sites.

      \begin{figure}[ht]
\center{\epsfig{figure=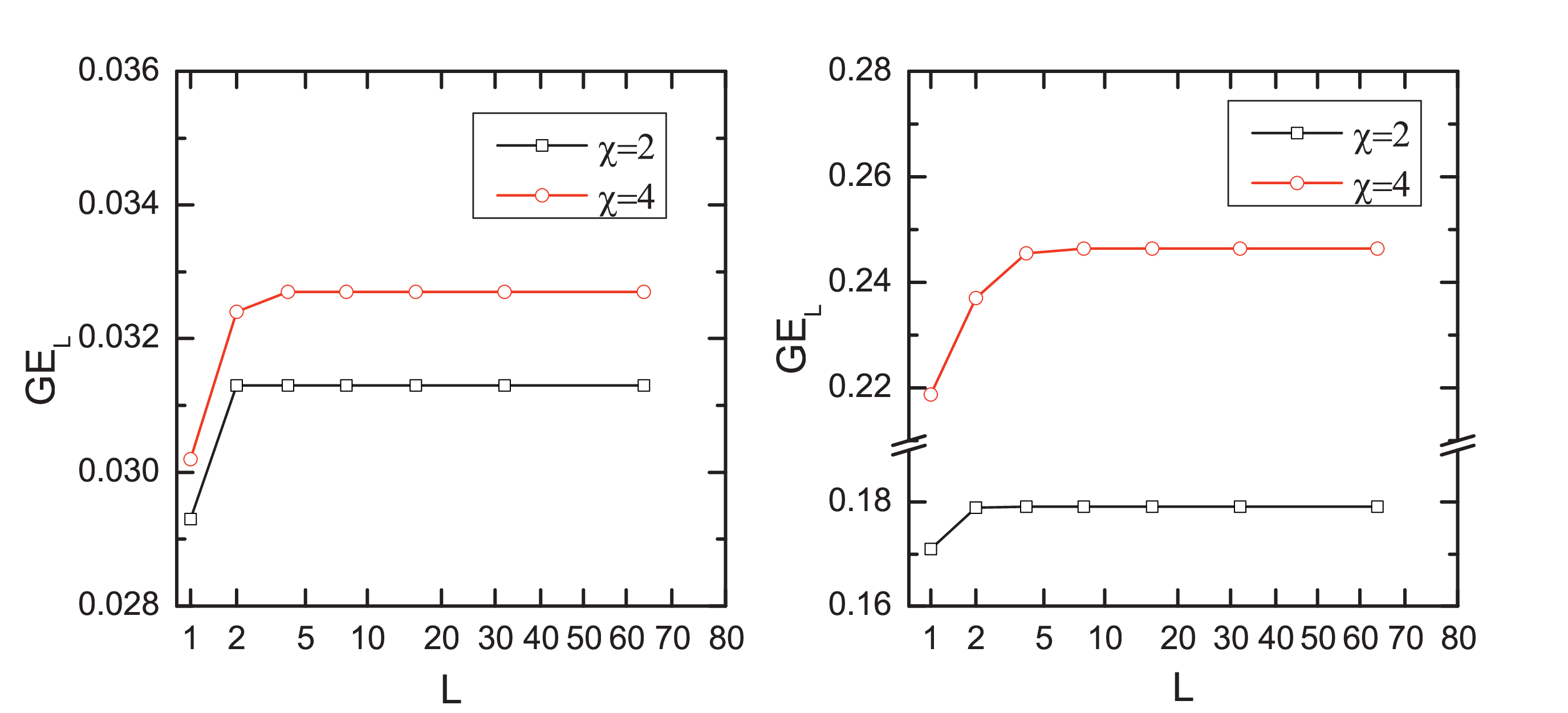,angle=0,width=16cm}}
   \caption{ The  GE of block size $L$ for spin 1 XXZ chain with the fixed $J_z=2.59$ and the quantum critical point at $D\simeq 2.3$.  We adopt the MPSs of $\chi=2,4$ for the ground state. (Left) Off the critical point, the block GE saturates around $L=4$. (Right) Very near the critical point, the block GE saturates around $L\simeq\chi^2$, which can be seen more clearly from the numerical data.}
   \label{scaling}
\end{figure}

      On the other hand, we like to see if the other entanglement measures will also obey the scaling law \eq{chis} and the  entropy bound \eq{SLb}.   We then evaluate the block GE for the spin 1 XXZ chain. Again, the direct computation of block GE is formidable because of the exponential growth of physical dimension for the trial separable states. Instead, one can adopt the method of quantum state RG transformation. In this way, one can still use one-site trial separable states but need to optimize the two-site unitary transformation for the quantum state RG transformation. This optimization only involves a vector of $\chi$ components, and it is numerically tractable if $\chi$ is not very large \footnote{In \cite{gemrg1,gemrg2,gemrg3} some analytical results was obtained for $\chi=2$ MPS of AKLT model. }.   Our result for the block GE of the spin 1 XXZ   chain with $\chi=2,4$ is plotted in Fig.  9. Again, we find that the results support the scaling law \eq{chis} and the entropy bound \eq{SLb}.

      For comparison, in Fig. 10 we also plot the  bipartite entanglement as a function of $\chi$, we see that it does  saturate off critical point, but the scaling law is $\xi_{\chi}\simeq \chi^{2.286}$ instead of \eq{chis}. It is not clear if the deviation is due to the truncation error of the approximate MPS or not.

       \begin{figure}[ht]
\center{\epsfig{figure=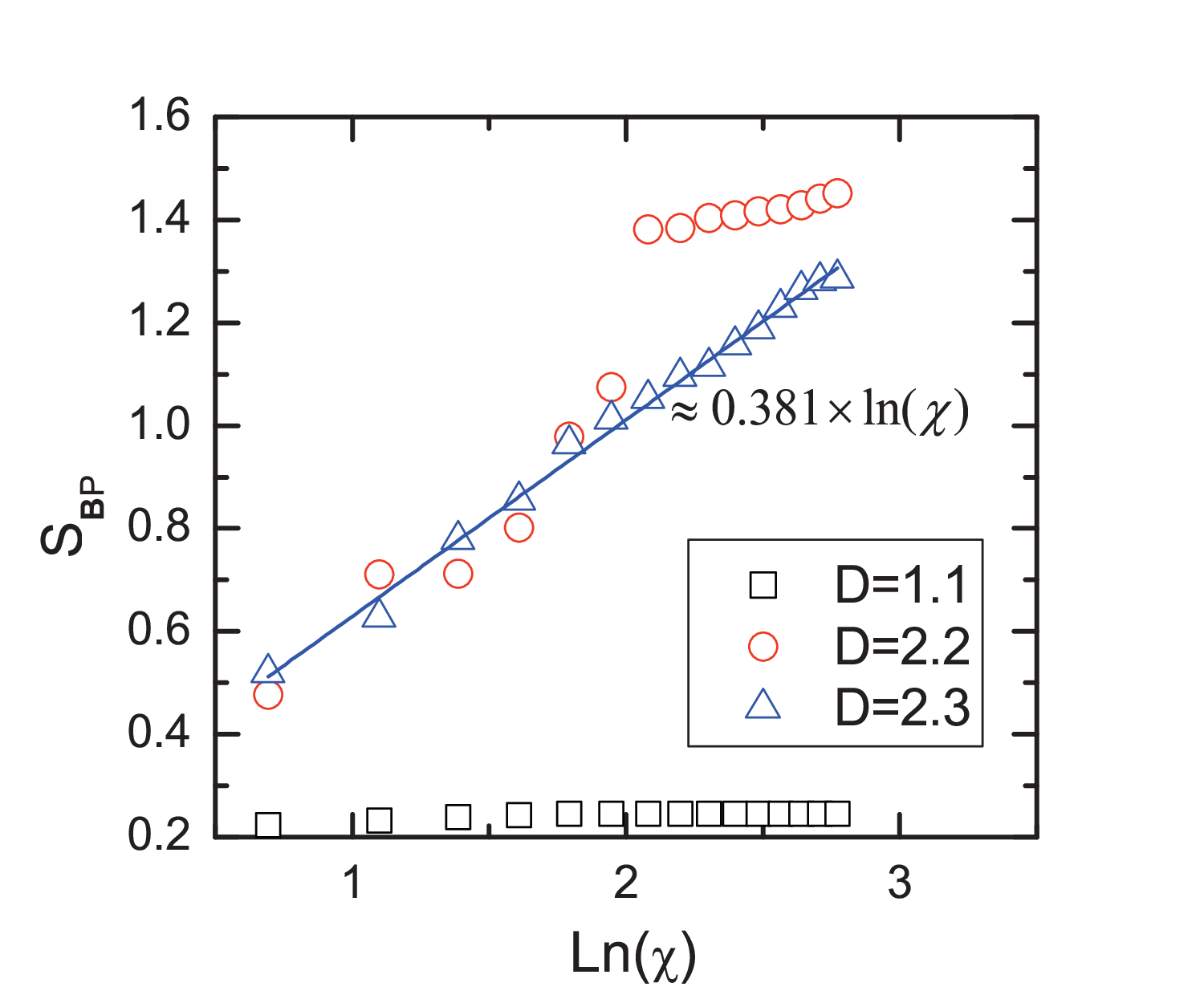,angle=0,width=8cm}}
   \caption{The scaling behavior of bipartite entanglement for spin 1 XXZ chain. The critical point is at $D\simeq 2.3$.}
   \label{scaling}
\end{figure}

\subsection{2D scaling via TPSs ?}

 We now aim to see if we can find the scaling law of the entanglement measure from the TPSs for 2D quantum spin systems.  The direct computation cannot work as in the 1D case, and we then wonder if the quantum state RG transformation works for our purpose or not.

   As in 1D case, to keep the physical dimension bounded after merging four neighboring sites into a new block by TRG, we then need to perform some appropriate SVD for the resulting tensor to obtain the non-local unitary transformation. Its physical dimension is bounded by $\chi^4$.  However, unlike the 1D case, the bond dimension will scale up exponentially after each merging, it implies that the bound on physical dimension will scale up accordingly. In fact, this is the same reason to introduce TRG method to calculate the double tensor trace by truncating the bond dimension, except now we do not contract out the physical indices when the performing quantum state RG transformations.

    To be more specific, let $A^s_{r,d,l,u}$ denote the tensor of a site for TPS with  $s$ the physical index, and $r,d,l,u$ the bond ones. After merging by usual TRG method, the resulting block tensor is denoted by $A^{(s_1, s_2, s_3, s_4)}_{r,d,l,u}$.  Note that $(s_1, s_2, s_3, s_4)$ is the combined physical index whose dimension is $d_s^4$, and the bond dimension is fixed as in the usual TRG method so that the bond indices $r, d, l, u$ remain intact.    Then, we perform the SVD so that $A^{(s_1, s_2, s_3, s_4)}_{r,d,l,u}\equiv \sum_{\delta=1}^{\min(d^4,\chi^4)}U^{\dag(s_1, s_2, s_3, s_4)}_\delta \lambda_{\delta} V^{\delta}_{r, d, l, u} $ to bound the physical dimension by $\chi^4$, or more precisely by  $ \min(d_s^4,\chi^4)$. After that, the non-local unitary operator $U^{\dag(s_1, s_2, s_3, s_4)}$ is used to RG transform the quantum state. Finally, we need to perform TRG again to keep $\chi$ fixed when evaluating the double tensor trace for the entanglement measure.

  \begin{figure}[ht]
\center{\epsfig{figure=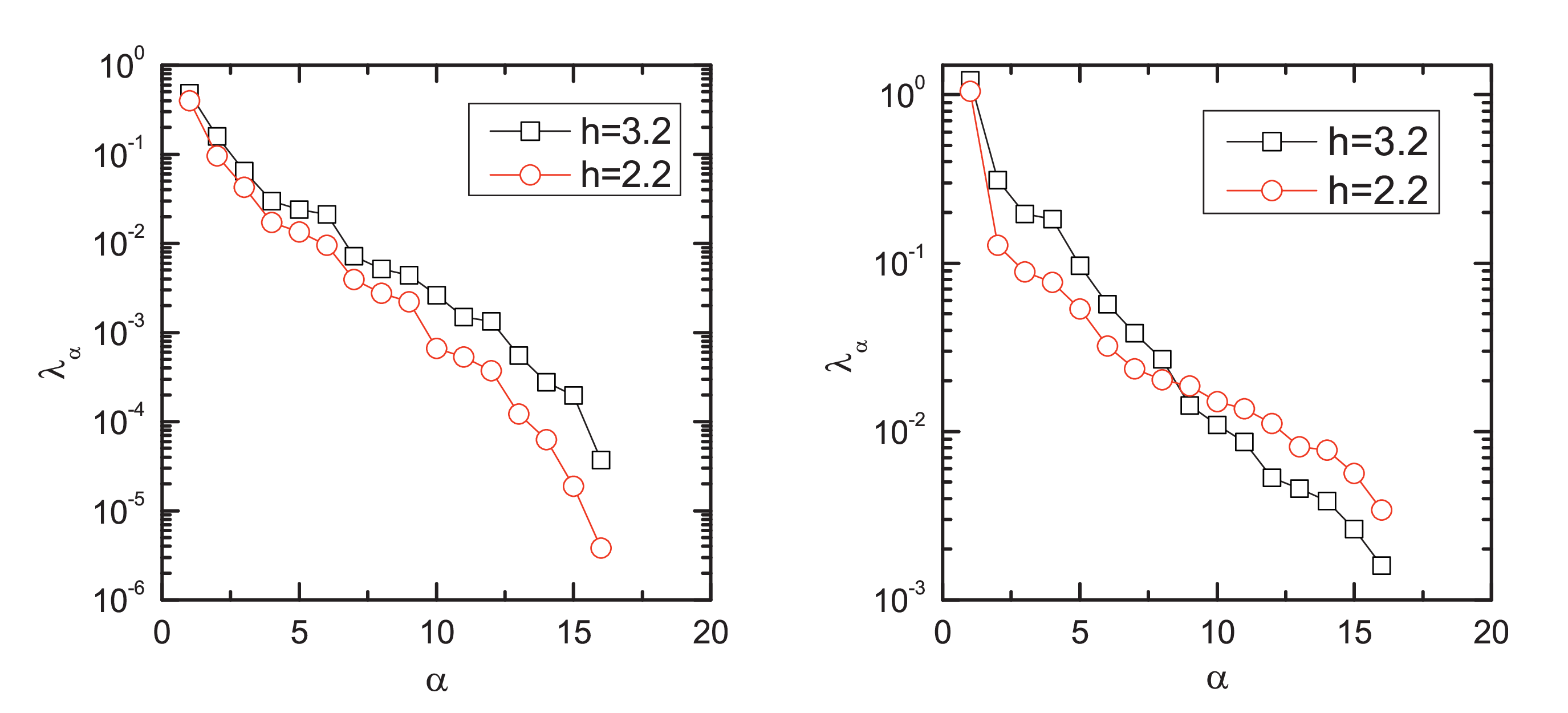,angle=0,width=16cm}}
   \caption{Spectra of singular values for initial bond dimension $\chi=4$. The index $\alpha$ is to number the singular values $\lambda_{\alpha}$'s. (Left) The spectrum of singular values for bond dimension after performing the TRG to merge the four neighboring sites. (Right) The spectrum of singular values for physical dimension after performing the SVD to find the non-local unitary transformation for quantum state RG. }
   \label{scaling}
\end{figure}

      Follow this prescription, we compute the block von Neumann entanglement entropy of the 2D transverse Ising model up to $\chi=4$.  However, we find that the resulting block entanglement decreases as the block size $L$ increases. We have also found that the resulting value of entanglement is not so sensitive to the value of physical dimension.  This failure may mean that we truncate the bond dimension too much in order to keep $\chi$ fixed when performing quantum state RG transformation, so that the entanglement is lost even when we increase the block size.  Of course, we could try to increase the $\chi$ value to keep more entanglement intact, but the computational cost for doing this is beyond what we can afford at this moment.  Instead, in Fig. 11 we plot the spectra of singular values after one step of quantum state RG transformation, which can indicate the loss of quantum entanglement due to the  truncation of  the bond and physical dimensions. It shows that we want to keep the singular values greater than $10^{-3}$, we should fix $\chi\simeq 10$ which, however,  is beyond our computation power.

 \*\\
 \section{Conclusion}

We have studied the global entanglement measures and bipartite entanglement entropy for various 1D and 2D quantum spin systems. This is achieved by the iTEBD and TRG algorithms.  Like the 1-tangle or concurrence, we find that both the global entanglement measure and the bipartite entanglement entropy are capable to characterize the quantum critical points as good as the order parameters. Moreover, they sometimes can capture the non-entangled states which are missed by the conventional order parameters. Especially, the bipartite entanglement entropy is far easier to compute in the MPSs/TPSs formalism than the most of the other entanglement measures. Interestingly, the bipartite entanglement entropy is always larger than the global entanglement measure for the cases studied here. It is hoped that one can understand if there is a universal class for the quantum entanglement measures, and the relation among the members of such a class.

  Besides, we also study the scaling behaviors of entanglement measures near the quantum critical point, which again shows some universal feature at least for the 1D spin systems. However, the method of the quantum state RG transformation fails to capture either the area law or the scaling law of the entanglement measures for 2D systems. It deserves more works to  elucidate the subtlety of truncating the bond and physical dimensions, and we hope to find a viable numerical algorithm in carrying out the scaling behaviors for 2D entanglement measures.

\section*{Acknowledgements}
  We thank Po-Chung Chen, Ming-Che Chang, Chao-Chun Huang, Li-Yi Hsu, Ying-Jer Kao, Keng-Shou Wu, Ling Wang, Ming-Fong Yang and I-Ching Yu for many helpful discussions. Especially, we thank Wo-Lung Lee for his generous support on the computer facility, and Tzu-Chieh Wei for his careful reading of our manuscript. We also thank the support of NCTS. This work is supported by Taiwan's NSC grant 097-2811-M-003-012.


\begin{thebibliography}{99}


\bibitem{qpt} S. Sachdev, ``Quantum Phase Transitions" (Cambridge University Press, Cambridge, England, 1999)


\bibitem{entan1}  T. J. Osborne and M. A. Nielsen, Phys. Rev. A \textbf{66}, 032110 (2002).

\bibitem{entan2}  A. Osterloh, L. Amico, G. Falci, and R. Fazio, Nature (London) \textbf{416}, 608 (2002).

\bibitem{entan3}  L. Amico, R. Fazio, A. Osterloh, and V. Vedral,  Rev. Mod. Phys. \textbf{80}, 517 (2008)




\bibitem{gem1} T.-C. Wei and P. M. Goldbart, Phys. Rev. A \textbf{68}, 042307 (2003).

\bibitem{gem2} T.-C. Wei, D. Das, S. Mukhopadyay, S. Vishveshwara, and P. M. Goldbart, Phys. Rev. A \textbf{71}, 060305 (2005).

\bibitem{gem3} Q.-Q. Shi, R. Orus, J. O. Fjaerestad, H.-Q. Zhou, arXiv:0901.2863 (2009).

\bibitem{gem4} R. Orus, T.-C. Wei, arXiv:0910.2488.

\bibitem{gemfd} H.-Q. Zhou, R. Or\'{u}s and G. Vidal, Phys. Rev. Lett. \textbf{100}, 080601 (2008).


\bibitem{dmrg} 	S. R. White,  Phys. Pev. Lett. \textbf{69}, 2863 (1992);  S. R. White,  Phys. Rev. B. \textbf{48}, 10345 (1993).


\bibitem{mps}	S. \"{O}stlund and S. Rommer,  Phys. Rev. Lett. \textbf{75}, 3537(1995);  S. \"{O}stlund and S. Rommer,  Phys. Rev. B. \textbf{55}, 2164(1996).

\bibitem{tebd}  G. Vidal,  Phys. Pev. Lett. \textbf{93}, 040502(2004); G. Vidal,  Phys. Rev. Lett. \textbf{91}, 147902(2003).

\bibitem{peps} 	F. Verstraete, J. I. Cirac,   arXiv:cond-mat/0407066;  V. Murg, F. Verstraete, and J. I. Cirac,   Phys. Rev. A. \textbf{75}, 033605(2007).

\bibitem{itebd} G. Vidal,  Phys. Rev. Lett. \textbf{98}, 070201(2007);	R. Or\'{u}s and G. Vidal,  Phys. Rev. B. \textbf{78}, 155117 (2008).



\bibitem{mera}	G. Vidal,  Phys. Rev. Lett. \textbf{99}, 220405(2007); G. Vidal,  Phys. Rev. Lett. \textbf{101}, 110501(2008).

\bibitem{ipeps} J. Jordan, R. Or\'{u}s, G. Vidal, F. Verstraete, and J. I. Cirac, Phys. Rev. Lett. \textbf{101}, 250602(2008)

\bibitem{trg1} M. Levin and C. P. Nave, Phys. Rev. Lett. \textbf{99}, 120601 (2007).

\bibitem{imt} H.-C. Jiang, Z.-Y. Weng, and T. Xiang, Phys. Rev. Lett. \textbf{101}, 090603 (2008);  Z. Y. Xie, H. C. Jiang, Q. N. Chen, Z. Y. Weng, T. Xiang, arXiv:0809.0182.

\bibitem{trg2} Z. C. Gu, M. Levin and X. G. Wen, Phys. Rev. B. \textbf{78}, 205116 (2008).

\bibitem{trg3} L. Wang, Y.J. Kao, A. W. Sandvik, arXiv:0901.0214




\bibitem{spin11}   H. J. Schulz,  Phys. Rev. B \textbf{34}, 6372 (1986)

\bibitem{spin12}  Y. Hatsugai and M. Kohmoto, Phys. Rev. B \textbf{44}, 11789 (1991)

\bibitem{spin13}   W. Chen, K. Hida, and B. C. Sanctuary,  Phys. Rev. B \textbf{67}, 104401 (2003).

\bibitem{2DIsing1} H. W. J. Bl\"{o}te and Y.  Deng, Phys. Rev. E \textbf{66}, 066110 (2002)



\bibitem{2DXYX1} T. Roscilde, P. Verrucchi, A. Fubini, S. Haas, and V. Tognetti, Phys. Rev. Lett. \textbf{94}, 147208 (2005).

\bibitem{2DXYX2}  B. Li, S.-H. Li, and H.-Q. Zhou, Phys. Rev. E \textbf{79}, 060101(R) (2009).

\bibitem{2DXXZ1}  S. Yunoki, Phys. Rev. B \textbf{65}, 092402 (2002).

\bibitem{2DXXZ2}  M. Kohno and M. Takahashi, Phys. Rev. B \textbf{56}, 3212 (1997).

\bibitem{2DXXZ3} P.-C. Chen, C.-Y. Lai, and  M.-F. Yang, arXiv:0905.4110.





\bibitem{qsrg} F. Verstraete, J. I. Cirac, J. I. Latorre, E. Rico, and M. M. Wolf, Phys. Rev. Lett. \textbf{94}, 140601 (2005).


\bibitem{gemrg1} R. Or\'{u}s, Phys. Rev. Lett. \textbf{100}, 130502 (2008).

\bibitem{gemrg2} T.-C. Wei, arXiv:0810.2564v2 (2009).

\bibitem{gemrg3} R. Or\'{u}s, Phys. Rev. A \textbf{78}, 062332 (2008).

\bibitem{gemrg4} A. Botero and B. Reznik,  arXiv:0708.3391 (2007).



\bibitem{Holzhey:1994we}
  C.~Holzhey, F.~Larsen and F.~Wilczek,
  Nucl.\ Phys.\  B {\bf 424}, 443 (1994)

\bibitem{entropyS1}  G. Vidal, J. I. Latorre, E. Rico, and A. Kitaev, Phys. Rev. Lett. \textbf{90}, 227902 (2003)

\bibitem{entropyS2} J. I. Latorre, E. Rico, and G. Vidal, Quantum Inf. Comput. \textbf{4}, 48 (2004)

\bibitem{entropyS3} L. Tagliacozzo, T. R. de Oliveira, S. Iblisdir, and J. I. Latorre,  Phys. Rev. B \textbf{78}, 024410 (2008).

\bibitem{entropyS4} F. Pollmann, S. Mukerjee, A. M. Turner, and J. E. Moore, Phys. Rev. Lett. \textbf{102}, 255701 (2009).

\bibitem{entropyS5}  J. I. Latorre, A. Riera, arXiv:0906.1499 (2009).

\bibitem{1DXY}
M. Henkel, ``Conformal Invariance and Critical Phenomena"  (Springer, Berlin, 1999)




\end{thebibliography}
\end{document}